\magnification=\magstep1
\font\bfv=cmbx10 scaled \magstep1
\def\ri{{\rm i}}%
\def\rd{{\rm d}}%
\def\izbaci#1{}%
\def\onetitle#1{%
\par\nobreak
\vskip\baselineskip
\vfil\penalty1000\vfilneg
\vbox{\leftline{#1}
\vskip0.75\baselineskip}\par\nobreak
}%
{
\null\vskip\baselineskip
{\leftskip=20pt plus 1fill\rightskip=20pt plus 1fill
\parindent=0pt\parfillskip=0pt
\bfv NON-CLASSICAL BEHAVIOR OF ATOMS IN AN 
INTERFEROMETER\par}
 \vskip.5cm

 \centerline{Lep\v sa Vu\v skovi\'c}
 \centerline{\it Old Dominion University, Department of Physics,}
 \centerline{\it 4600 Elkhorn Avenue, Norfolk, VA 23529}
 \vskip.3cm
 \centerline{Du\v san Arsenovi\'c and Mirjana Bo\v zi\'c}
 \centerline{\it Institute of Physics, P.O.Box 57,}
 \centerline{\it 11000 Belgrade, Yugoslavia}
 \vskip1truecm
 \centerline{\bf Abstract}
 \vskip0.25\baselineskip
{\narrower


We have studied the properties of the non-classical behavior of atoms
in a double-slit interferometer. An indication of this behavior for
metastable helium was reported by Kurtsiefer, Pfau and Mlynek [Nature
{\bf 386}, 150 (1997)] showing distinctive negative values of the
Wigner function, which was reconstructed from the measured diffraction
data. Our approach to explain this non-classical behavior is based on
the de Broglie-Bohm-Vigier-Selleri understanding of the wave-particle
duality and compatible statistical interpretation of the atomic wave
function.  It follows from the results that the atomic motion is
non-classical because it does not obey the laws of classical mechanics.
However, there is no evidence that this atomic behavior violates the
classical probability law of the addition of probabilities.
 
\par}
 \vskip0pt plus 1.5fill PACS number: 03.65.Bz, 03.75.Dg, 03.75.$^*$


{\tolerance=10000
Corresponding author: Mirjana Bo\v zi\'c,
{\it Institute of Physics, P.O.Box 57,}
{\it 11000 Belgrade, Yugoslavia,}
Tel: 381-11-3160260; Fax: 381-11-3162190; e-mail: \hbox{\tt bozic@phy.bg.ac.yu}

} \eject
}
 \leftline{\bf I. Introduction}
 \vskip0.75\baselineskip

The wave function $\psi(x,\,t)$ of the transverse motion of an atom in
an atom interferometer is
a linear superposition of states with maxima at two spatially separated
locations.  These states lead to negative values in Wigner's function
$W(x,\,p_x,\,t)$ which is the quasi-probability distribution of
coordinate $x$ and momentum $p_x$. The negative values of
$W(x,\,p_x,\,0)$, reconstructed from measured and evaluated space
distribution, were interpreted as a  signature of  the highly
non-classical behavior of atoms in the atom interferometer [1-3].

In this paper we study the properties and the cause of this
non-classical behavior, using the compatible statistical interpretation
(CSI) of a wave function [4-6]. The aim of this study is to clarify the
meaning of the notion ``non-classical motion (behavior)'' of atoms in a
double-slit interferometer. In our opinion, it is necessary to
distinguish clearly two aspects of the notion ``non-classical motion
(behavior)". It may denote a motion (behavior) which does not obey the
laws of classical mechanics and/or a motion which does not obey the
classical probability laws, in particular the classical law of the
addition of probabilities.

We use CSI because the wave and corpuscular features of a wave function
are incorporated in a consistent manner into the basic statistical
quantity of CSI, which is the de Broglian probability density
$P(x,\,p_x,\,t)$. $P(x,\,p_x,\,t)$ is the probability density for a
particle, which is in the quantum state $\psi(x,\,t)$, to have a
momentum $p_x$ and to be at $x$ at the time $t$ [5].  $P(x,\,p_x,\,t)$
satisfies both marginal conditions imposed by Wigner upon any joint
probability distribution in phase space and it is always positive [6].

The coherence and the characteristic modulation of the momentum
distribution found by Kaiser {\it et al.} [7] at the exit of a neutron
interferometer was explained by Bo\v zi\'c and Mari\'c [5], based on
the $P(x,p_x,t)$ function. Bo\v zi\'c and Arsenovi\'c [10] compared the
explanation of the same effect, based on Wigner's function and given by
Lerner, Rauch, and Suda [8] and Suda [9], with an explanation based on
the de Broglian probability density [5].  In this paper we use an
analogous comparison of the time dependent Wigner's function to the de
Broglian probability density.  These are two different distributions in
phase space, associated with the same wave function of the atomic
transverse degree of freedom in the interferometer.

In the following sections of this paper we summarize quantum properties
of atomic motion in an interferometer. In Sec. II is written the
solution of Schr\"odinger's equation for an interferometer in the form
of the Fresnel-Kirchhoff diffraction integral, while in Sec. III is
derived the time dependent wave function $\psi(x,\,t)$ of the
transverse motion and, for a chosen set of parameters, graphs of the
function $\vert\psi(x,t)\vert^2$ are presented. The transverse momentum
distribution in the state $\psi(x,\,t)$ is evaluated and presented
graphically in Sec. IV. The de Broglian probability density and
Wigner's function in the state $\psi(x,\,t)$ are evaluated and
graphically presented in Sects. V and VI, respectively. The comparison
between these two probability distributions is also given in Sec. VI,
while in Sec. VII are derived concluding remarks about the properties
of non-classical behavior of atoms in an interferometer.

\onetitle{
 \leftline{\bf II. The application of the Fresnel-Kirchhoff
diffraction formula}}

We want to determine the wave function of the transverse motion of an
atom which travels with velocity
$\vec v=v\vec i_y$ through region I (see Fig.~1), towards the slits and
is then sent through the slits to region II. For this reason we shall
determine in this section a
stationary solution of the time-dependent two dimensional Schr\"odinger
equation
$$
-{\hbar^2\over2m}\left({\partial^2\over\partial
x^2}+{\partial^2\over\partial
y^2}\right)\Psi(x,\,y,\,t)=\ri\hbar{\partial\over\partial
t}\Psi(x,\,y,\,t).\eqno(1)
$$
The stationary solution of Eq. $(1)$ has the form
$$
\Psi(x,\,y,\,t)=e^{-\ri\omega t}\Phi(x,\,y),\eqno(2)
$$
where $\hbar\omega = p^2 /2m$ and  $p  =  mv  =  \hbar k$. The space
dependent function  $\Phi(x,\,y)$ satisfies the equation
$$
-{\hbar^2\over2m}\left({\partial^2\over\partial
x^2}+{\partial^2\over\partial
y^2}\right)\Phi(x,\,y)=\hbar\omega\Phi(x,\,y).\eqno(3)
$$
The solution of Eq.~$(3)$ in region I is a spherical wave
$$
\Phi(P^\prime)=\Phi(x^\prime,\,y^\prime)={A{e^{\ri
kr}}^\prime\over r^\prime},\eqno(4)
$$
\noindent where $A$ is a constant and $r^\prime$ is the distance
(Fig.~1) from the source ($P_0$) to the point
$P^\prime=(x^\prime,\,y^\prime)$ in region I.  The spherical wave at
the slit points ($x^\prime,\,y^\prime=0$) may be  approximated by a
plane wave, since the distance $a$  of  the  double-slit screen from
the source $P_0$ is very large compared to  the  width  of  the slits.
Consequently, without a loss of generality, for $\Phi(x^\prime,y^\prime
= 0)$ at the border of region I we may choose the function
$$
\phi_1(x^\prime,0)=\cases{1/\sqrt{\delta},& $-{\Delta\over2}\ge
x^\prime\ge -{\Delta\over2}-\delta$\cr
0,&all other values of
$x^\prime$\cr}\eqno(5a)
$$
for one open slit, and the function
 $$
 \phi_2(x^\prime,0)=\cases{1/\sqrt{2\delta},& $-{\Delta\over2}\ge
 x^\prime\ge -{\Delta\over2}-\delta$\cr
 1/\sqrt{2\delta},
 &${\Delta\over2}+\delta\ge x^\prime\ge {\Delta\over2}$\cr
 0,&all other values of $x^\prime$\cr}\eqno(5b)
$$
for two open slits. This means that in region II the solution of Eq.
$(3)$ is given by the formula of the Fresnel-Kirchhoff diffraction [11]
$$
\Phi(x,\,y)=-{\ri A\over2\lambda}{e^{\ri ka}\over a}\int_{\cal
A}\rd x^\prime{e^{\ri ks}\over s}[1+\cos\chi],\eqno(6)
$$
where $s=\sqrt{y^2+(x^\prime-x)^2}$, $\cos\chi=y/s$, $\lambda=2\pi/k$,
while ${\cal A}=\{x^\prime;\,-(\Delta/2)-\delta<x^\prime<-(\Delta/2)\}$
when the lower slit is open and upper slit is closed, and ${\cal
A}=\{x^\prime;\,(\Delta/2)<x^\prime<(\Delta/2)+\delta$ or
$-(\Delta/2)-\delta<x^\prime<-(\Delta/2)\}$ when the two slits are
open. The constant $A$ will be chosen from the normalization condition.

The spatial distribution of the transverse degree of freedom as a
function of evolution time was investigated in a double slit experiment
[1,2] with metastable helium atoms. We will apply the Fresnel-Kirchhoff
diffraction formula to analyze the experiment of Kurtsiefer, Pfau, and
Mlynek [1]. A diagram of the apparatus is shown in Fig.~2.  Atoms are
emitted from a gas-discharge source operating in the pulse operation
mode.  The beam is collimated by a $5-\mu$m-wide slit and then is sent
through a double-slit structure with a slit separation
$\Delta+\delta=8\,\mu{\rm m}$ and an opening $\delta=1\,\mu{\rm m}$.
The atoms then propagate for a distance $d$ to a time- and
space-resolving detector. Atom beam velocities lie between $1000$ and
$3000\,{\rm ms^{-1}}$. We shall use the parameters of this experimental
arrangement for the following calculations.

 \vskip\baselineskip
 \leftline{\bf III. Time-dependent wave
function of the transverse motion}
 \vskip0.75\baselineskip
Assuming that the motion of an atom along the y-axis can be treated
classically and that the transverse motion is quantized, one may use
the relation $y = v t$ and determine the time dependent function of the
transverse motion from the function  $\Phi(x,\,y)$, by  the following
definition
$$
\Phi(x,\,y)=\Phi(x,\,vt)\equiv\psi(x,\,t).\eqno(7)
$$
The graphs of the function
$\vert\Phi(x,\,y)\vert^2\equiv\vert\psi(x,\,t)\vert^2$ for
$k=4\pi\cdot10^{10}\,{\rm m}^{-1}$ and for the chosen  set  of  values
of  the  coordinate  $y$ ($t=my/\hbar k$)  are presented in Figs.~3 and
~4.

Very close to the slit on the single-slit graphs (Fig.~3) we see the
minima of the wave function for $x = x_c$, where $x_c = - 4$ $\mu {\rm
m}$ is the coordinate of the slit center. But, with increasing $y$, the
maximum is present at $x = x_c$ for all $y$. This maximum becomes wider
and wider with increasing $y$.

On double-slit graphs (Fig.~4) we clearly see that near the slits the
wave function consists of two widely separated Gaussians on which small
oscillations are superimposed. With increasing distance from the slits
the Gaussian-like maxima spread and start to overlap, so that the third
maximum with superimposed oscillations start to develop. This region of
$y$ corresponds to Fresnel diffraction. With further increase of $y$
($t$), distinct equally spaced oscillations develop, which correspond
to the Fraunhofer diffraction limit.

 \vskip\baselineskip
 \leftline{\bf IV. The transverse-momentum distribution}
 \vskip0.75\baselineskip

The time dependent function defined by Eq. (7) should be a solution of
the one-di\-men\-si\-o\-nal time-dependent Schrodinger's equation.
Therefore, we may assume that it can be written in the form
$$
\psi(x,t)={1\over\sqrt{2\pi\hbar}} \int_{-\infty}^\infty
c(p_x)e^{ip_{x}x/\hbar}e^{-i\omega_{x}t}dp_x={1\over\sqrt{2\pi}}
\int_{-\infty}^\infty c^\prime(k_x)e^{ik_{x}x}e^{-i\omega_{x}t}dk_x,\eqno(8)
$$
where $\int_{-\infty}^\infty\vert c(p_x)\vert^2dp_x=\int_{-\infty}^\infty\vert
c^\prime(k_x)\vert^2dk_x=1$, $p_x = \hbar k_x$,
$c^\prime(k_x)=\sqrt{\hbar} c(p_x)$ and $\hbar\omega_x=p_x^2/2m$.
From Eq. (8) we may determine the transverse-momentum distribution $\vert
c(p_x)\vert^2=\vert c^\prime(k_x)\vert^2/\hbar$ in the state
$\psi(x,t)$. At first, one determines
$$
C(k_x,t)\equiv{1\over\sqrt{2\pi}}\int^\infty_{-\infty}\psi(x,t)e^{-
ik_{x}x}dx\eqno(9)
$$
by performing the Fourier-transform of the function $\psi(x,t)$,
defined by Eq. $(7)$, taking $t$ as a parameter. If Eq. (8) is valid,
then it should be
$$
C(k_x,t) = c^\prime(k_x)e^{-i\omega_{x}t}.\eqno(10)
$$
Consequently, 
$$
\vert c^\prime(k_x)\vert^2 = \vert C(k_x,t)\vert^2.\eqno(11)
$$
The graph of $\vert c^\prime(k_x)\vert^2 = \hbar\vert c(p_x)\vert^2$
for one slit is given in Fig. 5a and for two slits in Fig. 5b.

Our numerical calculation for various values of $t$, show that $\vert
c^\prime(k_x)\vert^2$ is independent of $t$. This fact justifies the
assumptions of Eqs. (7) and (8) as well as the statement of Kurtsiefer,
Pfau, and Mlynek [1] that the longitudinal motion of the atoms at
velocities $v$ of several thousand meters per second can be treated
completely classically.

We compared also the transverse momentum distribution $\vert
c^\prime(k_x)\vert^2$ (evaluated as described above and presented in 
Fig.~4) with the absolute value square of the Fourier transform

$$
F_i(k_x)={1\over\sqrt{2\pi}}\int^\infty_{-\infty}\phi_i(x^\prime,0)
e^{-ik_{x}x^{\prime}}dx^\prime\eqno(12)
$$
of the function $\phi_i(x^\prime,\,0), \ i = 1,2$. After the evaluation
of the latter integral one finds

$$
\eqalignno{
F_1(k_x)&={ie^{ik_{x}\Delta/2}\over
k_x\sqrt{2\pi\delta}}\{1-e^{ik_{x}\delta}\};&\cr 
\vert
F_1(k_x)\vert^2 &= {2\sin^2(k_x\delta/2)\over\pi\delta k_x^2}&(13)\cr}
$$
and
$$
\eqalignno{
F_2(k_x)&= {2\over
k_x\sqrt{\pi\delta}}\sin{k_x\delta\over2}\cos{k_x(\Delta+\delta)\over2};\cr
\vert F_2(k_x)\vert^2 &={4\over
k_x^2\pi\delta}\sin^2{k_x\delta\over2}\cos^2
{k_x(\Delta+\delta)\over2}.&(14)\cr}
$$
We found that $\vert c^\prime(k_x)\vert^2$ for one slit is practically
identical to $\vert F_1(k_x)\vert^2$ and that $\vert
c^\prime(k_x)\vert^2$ for two slits is practically identical to $\vert
F_2(k_x)\vert^2$.

By comparing the spatial distributions for one and two slits shown in
Figs.~3 and ~4, one must conclude that the presence of the second slit
influences the motion of each atom, independent of the slit through
which it has passed to region II (see Fig. 1).  This influence is also
very well seen by comparing the momentum distributions for one and two
slits, presented in Fig.~5. Certain values of the particle's transverse
momentum, which are allowed with one slit, are not allowed when both
slits are open.  This fact is also a signature of a non-classical
atomic behavior that can be understood in a similar way to the
quantization of the electronic orbits in atom based on de Broglie's
wavelength. It appears that the atomic matter wave excludes certain
values of transverse momentum and favors others, which is an evident
quantum effect.

 \vskip\baselineskip
 \leftline{\bf V. The De Broglian probability density}
 \vskip0.75\baselineskip

According to the CSI of a wave function, in an ensemble of particles in
a pure state presented by Eq. $(8)$, different particles may have
different momenta. Recall that the probability density of $p_x$ is
$\vert c(p_x)\vert^2$. However, each particle is surrounded by the same
wave [5]. In other words, a particle and a wave are two different, but
compatible, entities.

The de Broglian probability density, $P(x,p_x,t)$, of a particle in the
quantum state $\psi(x,t)$ is the probability density for the particle
to have a momentum $p_x$ and to be at a position $x$ at time $t$
[5,10],
$$
 P(x,p_x,t)= \vert\psi(x,t)\vert^2\vert c(p_x)\vert^2 =
 P^\prime(x,k_x,t)/\hbar=\vert\psi(x,t)\vert^2\vert
 c^\prime(k_x)\vert^2/\hbar.\eqno(15)
 $$
 $P(x,p_x,t)$ satisfies both marginal conditions
 $$
 \int P(x,p_x,t)dp_x=\vert\psi(x,t)\vert^2,\eqno(16)
 $$
 $$
 \int P(x,p_x,t)dx=\vert c(p_x)\vert^2\eqno(17)
 $$
imposed by Wigner upon a joint probability distribution in phase space
[12]. For operators having the form $F(\hat x,\hat p_x)=F_1(\hat
x)+F_2(\hat p_x)$, the probability density $P(x,p_x,t)$ satisfies
Wigner's condition that the quantum mechanical average value of an
operator is equal to the classical average value of the corresponding
classical function.

Both  $P(x,p_x,t)$ and  Wigner's function are determined by the state
$\psi(x,t)$. Unlike the Wigner function,  $P(x,p_x,t)$ is always
positive.  Despite this fact, $P(x,p_x,t)$ also reflects the
non-classical behavior of atoms in the state $\psi(x,t)$.

Since the simultaneous measurement of a coordinate and momentum is not
possible, $P(x,p_x,t)$ cannot be measured in a single experiment.
However, one could experimentally determine the probability density of
a coordinate $x$ and momentum $p_x$ in the state $\psi(x,t)$, {\it
i.e.} $P(x,p_x,t)$, by measuring separately the distributions
$\vert\psi(x,t)\vert^2$ and $\vert c(p_x)\vert^2$. These distributions
reflect the non-classical behavior, as pointed out in the previous
section.

 \vskip\baselineskip
 \leftline{\bf VI. The De Broglian probability distribution and Wigner's function}
 \vskip0.75\baselineskip

In Figs.~6 and ~7 we present the graphs of the de Broglian probability
density of a coordinate $x$ and transverse momentum $p_x$,
$P(x,\,p_x,\,t)$, for $y=120$ mm $(t=y/v=6.01\times10^{-5}{\rm s})$ and
$y=240$ mm $(t=y/v=12.02\times10^{-5} {\rm s})$. For the same values of
$y$ we present in Figs.~8 and ~9 the plots of the Wigner distribution
function, evaluated from the definition [12,13] expression
$$
W(x,\,p_x,\,t)={1\over\hbar\pi}\int\rd\tilde xe^{2\ri p_x\tilde
x/\hbar}\psi^*(x+\tilde x,\,t)\psi(x-\tilde
x,\,t)=W^\prime(x,\,k_x,\,t)/\hbar.\eqno(18)
$$

It is clear from Figs.~6-9 that $W(x,\,p_x,\,t)$ and $P(x,\,p_x,\,t)$
are very different functions. Consequently, from their forms and
properties are derived different interpretations of the behavior of
quantum particles. It was shown by Janicke and Wilkens [14] and
Kurtsiefer, Pfau, and Mlynek [1] that Wigner's function
$W(x,\,p_x,\,0)$ may be reconstructed from evaluated and measured
values of $\vert\psi(x,\,t)\vert^2$ for various values of $t$. The
negative values of $W(x,\,p_x,\,0)$ were interpreted as a signature of
an atom's non-classical behavior. These negative values are also
associated with the requirement of Heisenberg's uncertainty
relationship that a quantum mechanical particle has to be described by
an area of uncertainty in phase space no smaller than $\Delta x\Delta
p_x=\hbar/2$ [3]. The authors pointed out that the negative values
reflect the impossibility of joint measurement of position and
momentum.

However, we interpret de Broglian probability density, presented in
Figs. ~6 and ~7, as an objective probability density of particle
coordinate and momentum. The impossibility of simultaneous measurements
of a particle's $x$ and $p_x$ does not forbid us from assuming that
their joint distribution objectively exists. The important fact is that
this assumption does not lead to any contradiction with the facts
derived from measurable distributions. One can see that this joint
probability density is consistent with the measurable probability
density of position and the measurable probability density of momentum.
For example, for values of  $\tilde p_x$ for which $\vert c(\tilde
p_x)\vert^2=0$, the joint distribution $P(x,\,\tilde p_x,\,t)$ is also
equal to zero. Thus, if there is no particle with a certain value of
momentum $\tilde p_x$, this value can not be found {\bf anywhere}
during the measurement of momentum. Similar reasoning is valid for
space points $\tilde x$ in which $\vert\psi(\tilde x,\,t)\vert^2=0$,
since $P(\tilde x,p_x,t)$ is also equal to zero in these space points
for any value of momentum. Therefore, at a point $\tilde x$ {\bf no}
particle will be detected in the experiment.

One can see in Figs. 8 and 9 that Wigner's function $W(x,p_x,t)$ may
take values different from zero at the points $\tilde x$ and $\tilde p$
in which either $\psi(\tilde x,\,t)=0$ or $c(\tilde p_x)=0$. Despite
this property, inconsistent with a notion of a joint probability, the
Wigner function satisfies the marginal
conditions stated by Eqs.  (16) and (17). It is well known that
Wigner's function may assume negative values, even though it is a joint
probability distribution by definition.  Because of this, it is
possible to satisfy Eqs. (16) and (17).  Thus, two different
properties of Wigner's function, inconsistent with a notion of a joint
probability, cancel each other and make it
possible to satisfy two marginal conditions. This is clearly seen by
comparing results presented in Figs. 6,7 and 8,9. We note in Fig. 8,9
the negative peaks in the $x$-dependence of the Wigner function for
those values $\tilde p_x$ of momentum for which $\vert c(\tilde
p_x)\vert^2=0$.

 \vskip\baselineskip
 \leftline{\bf VII. Conclusion}
 \vskip0.75\baselineskip

The properties and cause of non-classical behavior of atoms in the
atomic interferometer are studied using the stationary solution
$\Phi(x,\,y)$ of the two-dimensional Schrodinger's equation. The
solution was written in the form of the Fresnel-Kirchhoff diffraction
integral. The time dependent wave function of the transverse motion was
derived from the Fresnel-Kirchhoff diffraction integral, using the
relation $\Phi(x,\,y)=\Phi(x,\,vt)\equiv\psi(x,\,t)$, where $v$ is the
initial longitudinal atomic velocity.  The latter relation was used in
Refs. [1,2], where it was justified by the experimental facts,
suggesting that the longitudinal atomic motion was classical and that
the transverse motion was quantum. We determined the transverse
momentum distribution in the state $\psi(x,\,t)$, by evaluating its
Fourier transform.

We calculated $\vert\psi(x,t)\vert^2$ for one and two-slit
interferometers, and presented results in Figs. 3 and 4. From the data
one can see that the evolution (and spreading) of waves from different
single slits and their interference (overlap) determine Fresnel's and
Fraunhofer's regimes and the transition from the former to the latter.
We conclude that this spatial distribution, which reflects the
non-classical atomic motion, is due to a real atomic wave that is
associated with each atom and that influences its motion.  The obstacle
in front of the incoming atoms determines the concrete form of this
influence.

Our results show that the de Broglie [15], the Bohm and Vigier [16]
and the Selleri [17] understandings of wave-particle duality is
applicable to the explanation of the non-classical motion of atoms in
an interferometer. The conclusion that the motion is non-classical
means that it is different from the motion of a particle which obeys
the laws of classical mechanics.  This difference is due to the fact
that with a classical particle no wave is associated, whereas the atom
is accompanied by its wave. This conclusion follows from the measured
atomic distribution [1,2], and its theoretical explanation, based on
the particle's wave function, in this and in the previous works
[1,2,18]. Therefore, the application of methods for determination of
the amplitude and phase structure of the atomic wave field, like the
method of Raymer, Beck, and McAlister [19] would be of great
importance.
 
However, neither from the measured space distribution [1,2], nor from
the transverse momentum distribution evaluated in this paper, does it
follow that the motion of atoms in the double-slit interferometer
violates the classical probability laws, especially the law of the
addition of probabilities. From the fact that Wigner's function,
associated with the state $\psi(x,\,t)$, takes a negative value if does
not follow that atomic motion violates the latter law. It rather seems
that the classical probability laws are satisfied and are consistent
with the de Broglie-Bohm-Vigier-Selleri [15-17] understanding of the
wave-particle duality and the compatible statistical interpretation of
a wave function [4-6].

To the best of our knowledge, the atomic transverse momentum
distribution in the interferometer has not been measured. However, as
pointed out in this paper, it is very important characteristic of the
quantum state. Experimental evidence of the transverse momentum
distribution would contribute a lot to our understanding of the quantum
nature of atomic motion.

\vskip\baselineskip
\leftline{\bf Acknowledgments}

We acknowledge the communication with Tilman Pfau, who provided details
of his experiment and commented an early version of the manuscript.

D. A. and M. B. acknowledge support by Ministry of science and
technology of Republic of Serbia under contract 01M01.
\vfill\eject
\null 
\leftline{\bf References }
\vskip0.75\baselineskip
\frenchspacing
\item{1.}Ch. Kurtsiefer, T. Pfau, and J. Mlynek, Nature 386 (1997) 150.
\item{2.}T. Pfau and Ch. Kurtsiefer, J. Mod. Opt. in press. 
\item{3.}D. Leibfried, T. Pfau, and C. Monroe, Physics Today, April 1998, 
p. 22.
\item{4.}M. Bo\v zi\'c and Z. Mari\'c, Phys. Lett. A 158 (1991) 33.
\item{5.}M. Bo\v zi\'c and Z. Mari\'c, Found.  Phys. 25 (1995) 159. 
\item{6.}M. Bo\v zi\'c and Z. Mari\'c, Found. Phys. 28 (1998) 415.
\item{7.} H. Kaiser, R. Clothier, S.A. Werner, H. Rauch, and H. Wolwitsch, 
Phys. Rev. 45 (1992) 31. 
\item{8.} P.B.Lerner, H. Rauch, and M. Suda, Phys. Rev. A 51 (1995) 3889. 
\item{9.} M. Suda, Quantum Semiclass. Opt. 7 (1995) 901.
\item{10.}M. Bo\v zi\'c  and  D.  Arsenovi\'c,  in  Epistemological  and 
Experimental
  Perspectives on Quantum Physics, eds. D.  Greenberger  et  al.   (Kluwer,
  Dordrecht, 1999) p. 225.
\item{11.}M. Born and E. Wolf, Principles of Optics (Pergamon Press, 
Oxford, 1965).
\item{12.}M. Hillery, R.F. O'Connell, M.O.Scully, and E.P. Wigner, 
Phys. Rep. 106 (1984) 121.
\item{13.}Y.S. Kim and M.E. Noz, Phase Space  Picture  of  Quantum  
Mechanics  (World Scientific, Singapore, 1991).
\item{14.}U. Janicke and M. Wilkens, J. Mod. Opt. 42 (1995) 2183.
\item{15.} L. de Broglie, The Current Interpretation of Wave
Mechanics, A Critical Study (Elsevier, Amsterdam, 1964).
\item{16.} D. Bohm and J.P. Vigier, Phys. Rev. 96 (1954) 208.
\item{17.} F. Selleri, Ann. Fond. L. de Broglie 7 (1982) 45.
\item{18.}M. Freyberger and W.P. Schleich, Nature 386 (1997) 121.
\item{19.}M.G. Raymer, M. Beck, and D.F. McAlister, Phys. Rev. Lett. 
72 (1994) 1137.
\vfill
\eject
\null
\vskip\baselineskip
\centerline{\bf Figure captions}
\vskip0.75\baselineskip
Figure 1. Illustration of the diffraction formula presented with Eq. (6).

Figure 2. Diagram of apparatus used in Ref. [1] to observe atomic
interference patterns.

Figure 3. The function
$\vert\psi(x,\,t)\vert^2\equiv\vert\Phi(x,\,y=vt)\vert^2$ for a single
slit evaluated from Eq. $(6)$ with  $A$ chosen, from the condition
$\int\vert\Phi(x,\,y=vt)\vert^2\rd x=1$ for a given $y$. Other
parameters are: $k=4\pi\cdot10^{10}\,{\rm m}^{-1}$, $v=\hbar
k/m=1995.58$ m/s, $m=6.64632\cdot10^{-27}$ kg is the mass of the Helium
atom.

Figure 4. The function
$\vert\psi(x,\,t)\vert^2\equiv\vert\Phi(x,\,y=vt)\vert^2$ for a
double-slit evaluated from Eq. $(6)$ with  $A$ chosen, from the
condition $\int\vert\Phi(x,\,y=vt)\vert^2\rd x=1$, for a given $y$.
Other parameters are: $k=4\pi\cdot10^{10}\,{\rm m}^{-1}$, $v=\hbar
k/m=1995.58$ m/s, $m=6.64632\cdot10^{-27}$ kg is the mass of the Helium
atom.

Figure 5. Momentum distribution $\vert c^\prime(k_x)\vert^2=\vert
c(p_x)\vert^2\cdot\hbar$ in the state $\psi(x,\,t)$ with parameters
given in the caption of Figs. 3 and 4. {\it a\/}) One open slit, {\it
b\/}) Two open slits.

Figure 6. The de Broglian probability density
$P^\prime(x,\,k_x,\,t)=\hbar P(x,\,p_x,\,t)$ in the single slit state
$\psi(x,\,t)$ with parameters given in the caption of Fig. 3.

Figure 7. The de Broglian probability density
$P^\prime(x,\,k_x,\,t)=\hbar P(x,\,p_x,\,t)$ in the double-slit state
$\psi(x,\,t)$ with parameters given in the caption of Fig. 4.

Figure 8. Wigner's function $W^\prime(x,\,k_x,\,t)$ associated with the
single slit state $\psi(x,\,t)$ and evaluated from $(18)$. Parameters
are given in the caption of Fig. 3.

Figure 9. Wigner's function $W^\prime(x,\,k_x,\,t)$ associated with the
double-slit state $\psi(x,\,t)$ and evaluated from $(18)$. Parameters
are given in the caption of Fig. 4.

\end